\documentclass[journal,10pt]{IEEEtran}

\usepackage{times,amsmath,amsthm,epsfig}
\usepackage{pgfplots}
\pgfplotsset{width=25em,compat=1.9}

\usepackage{pst-plot} 
\usepackage{pstricks-add}

\usepackage{amssymb}%
\usepackage{color,colortbl}

\usepackage{hyperref}
\usepackage{multirow}

\definecolor{navy}{RGB}{0,0,128}
\newcommand{\revised}[1]{{#1}} 

\definecolor{Gray}{gray}{0.9}

\makeatletter
\def\subsubsection{%
  \@startsection
    {subsubsection}                 
    {3}                             
    {\parindent}                    
    {3.5ex plus 1.5ex minus 1.5ex}  
    {0ex}    
    {\bf\normalsize}
}
\makeatother

\usepackage{quoting,xparse}

\NewDocumentCommand{\bywhom}{m}{
  {\nobreak\hfill\penalty50\hskip1em\null\nobreak
   \hfill\mbox{\normalfont(#1)}%
   \parfillskip=0pt \finalhyphendemerits=0 \par}%
}

\NewDocumentEnvironment{pquotation}{m}
  {\begin{quoting}[
     indentfirst=true,
     leftmargin=\parindent,
     rightmargin=\parindent]\itshape}
  {\bywhom{#1}\end{quoting}}

\title{Data Science for Engineers:\\  A Teaching Ecosystem}

\author{
    \IEEEauthorblockN{Felipe Tobar\IEEEauthorrefmark{1}, Felipe Bravo-Marquez\IEEEauthorrefmark{2}, Jocelyn Dunstan\IEEEauthorrefmark{1}, Joaquín Fontbona\IEEEauthorrefmark{1}, \\ Alejandro Maass\IEEEauthorrefmark{1}, Daniel Remenik\IEEEauthorrefmark{1}, Jorge F. Silva\IEEEauthorrefmark{3}
    } \\ \vspace{1em}
    \IEEEauthorblockA{\IEEEauthorrefmark{1}Center for Mathematical Modeling, University of Chile} \\
    \IEEEauthorblockA{\IEEEauthorrefmark{2}Department of Computer Science, University of Chile \& IMFD} \\
    \IEEEauthorblockA{\IEEEauthorrefmark{3}Department of Electrical Engineering, University of Chile} \\

}

\begin{document}
\maketitle

\begin{abstract}
We describe an ecosystem for teaching data science (DS) to engineers which blends theory, methods, and applications, developed at the Faculty of Physical and Mathematical Sciences, Universidad de Chile, \revised{over the last three years}. This initiative has been motivated by the increasing demand for DS qualifications both from academic and professional environments.  The ecosystem is distributed in a collaborative fashion across three departments in the above Faculty and includes postgraduate programmes, courses, professional diplomas, data repositories, laboratories, trainee programmes, and internships. By sharing our teaching principles and the innovative components of our approach to teaching DS, we hope our experience can be useful to those developing their own DS programmes and ecosystems. The open challenges and future plans for our ecosystem are also discussed at the end of the article.
\end{abstract}

\section{Introduction}
\label{sec:intro}
\subsection{Interdisciplinarity to embrace new challenges}

The taxonomy of academic branches reflects the necessities of a society in space and time and is, therefore, subject to both gradual and sudden changes, just as the  evolution of science \cite{kuhn2012structure}. Early universities dealt with subjects such as theology and natural philosophy; wars catalysed the teaching of engineering (both civil and military), while schools in rural areas have grown to focus on agricultural studies, and business schools have arisen near financial districts. This suggests that the division we impose over different branches of knowledge, and in particular of science, is, to a large extent, purely instrumental: it obeys our own necessities and not an evident or natural segmentation \cite{knowledge2010gibbons}. As the necessities and interests of societies change over time,  academic branches evolve both in depth and scope. This problem-driven reformulation promotes the creation of new relevant fields and academic branches. However, this does not always occur in a timely manner, but is instead a lengthy process whose timing often lags behind urgent societal demands.

In response to increasing complex practical needs and societal demands, approaches that rely on the interaction of existing, well-studied branches of knowledge have recently come into focus. This interaction between or among disciplines is what is referred to as  \emph{interdisciplinarity/multidisciplinarity} \cite{klein1990interdisciplinarity} and has become essential to address current challenges effectively.  Multidisciplary/interdisciplinary approaches are effective because the skills and abilities required to confront today's challenges are segregated across different disciplines due to the outdated taxonomy of knowledge that was once imposed under different conditions.

\subsection{The case of data science}
A contemporary instance of the above phenomenon is that of \emph{Data Science} \cite{donoho2017,tukey1992}. From the public sector, industry and academia, a number of agents are demanding (and also offering) solutions that are labelled as data science (DS) or related terms such as artificial intelligence (AI), machine learning (ML), data mining, and big data.
Briefly put, a DS task is one that involves some of the following stages: acquisition, curation, transmission, processing, analysis, interpretation and visualisation of some form of information content. The skills needed to address current challenges in DS are mainly those found in machine learning \cite{murphy2012machine},  mathematics (optimisation \cite{boyd2004convex},
probability and statistics \cite{gelman2013bayesian}), computer science \revised{(data mining, semantic web and database theory \cite{abiteboul1995foundations})}, electrical engineering (signal processing \cite{oppenheim1999discrete}, estimation and detection, information theory \cite{mackay2003information}, and control theory \cite{bertsekas1995dynamic}), operations research \cite{winston2004operations}, and high-performance/scientific computing \cite{oliphant2007python} among others. In addition to the numerous stages involved in DS, the boundaries between the stages are not clearly defined. The complexity of DS, therefore, calls for a holistic approach, where mathematics, computer science and engineering among other disciplines collectively devise new strategies to address contemporary challenges. This rationale has shown to be particularly effective when it comes to addressing open problems in DS research, and it therefore should be incorporated in the way DS is taught.

Since the various disciplines that constitute DS research and practice are rarely found together, students,  professionals and even academics have struggled to cobble together pedagogical sources for DS training. As a consequence of this demand, we have more recently witnessed a proliferation of academic/professional programmes on DS offered by departments of engineering, business, statistics or computer science. The significance of DS training and practice \cite{the2019dynamics} has even been identified by policy makers as a key component of  national strategies on artificial intelligence in \revised{the UK, France, Canada, the USA, China and India to name a few.}

\subsection{Data is the new oil}

In Chile, we believe that DS is instrumental to migrate from a natural-resource economy to a knowledge-based one. The Chilean economy, in particular, is largely based on the exploitation of natural resources such as copper, agriculture, forestry, and fishing. In fact, companies in the mining and agriculture sectors employ a limited (or no) highly-trained workforce \cite{Doner2020}. 
As other developing countries, Chile has been updating its strategy for sustaining productivity and directing its economy towards knowledge-based growth \cite{OECD2018} -- a process where data is undoubtedly the raw element \cite{claudio2017}. Two popular quotes reflect well the role that DS plays in knowledge-based economies: \begin{pquotation}{Clive Humby, \cite{dataoil}}
``Data is the new oil. Data is just like crude. It is valuable, but if unrefined, it cannot really be used.''
\end{pquotation}
\begin{pquotation}{Harvard Business Review, \cite{davenport2012data}}
``Data science is the sexiest job of the 21st century.''
\end{pquotation}
These quotes suggest that DS is key for the future and developing countries such as Chile, where  alternative paths to consolidate the economy are urgently needed, must recognise this opportunity. 

The need for DS is, however, not exclusive to the economy but applies to the more general concept of \emph{societal development} reaching into every realm. The wide-ranging importance of DS has been identified by the authorities in Chile too, where a discussion towards a national strategy on AI is currently underway\footnote{See \url{https://www.minciencia.gob.cl/politicaIA}.}, as in many developed societies. However, once the ability to produce data is in place, the next step towards building a knowledge-based society is to equip citizens with the tools to extract value from that data. Therefore, teaching DS should be a prime objective for societal development and, thus, a duty of the academia. 






\subsection{Scope of the paper and organisation}

The scope of DS’s reach and the challenges it presents demand more than a programme but rather an ecosystem to address its development within the university. This article presents a set of initiatives that encompass teaching DS to engineering students at undergraduate and graduate levels, research interns, project engineers, and professionals. The DS ecosystem described in the following sections is not exclusive to any programme or department in our engineering school, but rather exists as a collection of campus-wide resources, which can be extended as demand requires.  Our description builds on the teaching and applied research experiences of the authors over the last three years, both independently and collaboratively, with emphasis on blending theory and practice as required in modern, real-world DS challenges.

The remainder of the paper is organised as follows. Section \ref{sec:perspective} presents the setting of the DS ecosystem in relationship to our University and the focus of DS learning objectives. Section \ref{sec:components} describes the components of the currently existing DS ecosystem, such as graduate programmes, courses and internships. Section \ref{sec:innovations}, the core of this article, presents the innovative aspects of what has been developed thus far.  Section \ref{sec:for_professionals} focuses on continuing education resources, while Section \ref{sec:outreach} explains how DS can be  \emph{democratised} to a broader non-technical community via outreach. Lastly, Section \ref{ssec:open_c} outlines the open challenges of the presented ecosystem.

\section{Our perspective on teaching DS}
\label{sec:perspective}

\revised{
Academic and professional programmes in data science have flourished across the country. This is in line with the global trend, where the success of DS generates a demand for data scientists with educational institutions aiming to fulfil such demand. In a multidisciplinary fashion, available DS programmes adopt different perspectives stemming from  business, computing, and the natural or social sciences. In this competitive arena, DS programmes must establish a clear and focused objective; ours, in particular, is that of engineering. 
}

\revised{ 
We believe that research, teaching, and practice are heavily intertwined in DS: those at the forefront of DS research and practice are the best-equipped to teach it. With this idea in mind, our approach adopts an \emph{engineering perspective}, blending concepts from mathematics and the natural sciences, resources from computer science, and applications of general interest as in \cite{curr_guide_2017}. This perspective has been instrumental to equip students with the necessary scientific background while also exposing them to real-world engineering challenges that come from various sectors of human endeavours, including industry, science, and engineering.}

\revised{

\subsection{Our engineering school in context} 
\label{sub:our_units} 

The Faculty of Physical and Mathematical Sciences (FCFM is its Spanish acronym) at Universidad de Chile was established in 1842 and (as of 2021) hosts 12 departments and 10 research centres. The \emph{civil engineering} degree at FCFM, as in most universities in Chile, requires 11-12 terms  (depending on each specific specialty) of full-time study spanning 6 years. After this period, our graduates receive both a \emph{Bachelor of Science in Engineering} (BSc) completed after the fourth year and a \emph{Professional Engineer's Title} (PET) oriented to practical engineering duties. The last credits of the degree focus on elective courses, which can be industry- or research-oriented, and the thesis work. Critically, some students enrol in MSc programmes in parallel with their last undergraduate year. MSc programmes at FCFM are 2 years long, yet the joint BSc, PET and MSc degree can be completed in 7 years due to overlapping requirements. In addition to the academic programmes, FCFM also offers professional diplomas on different engineering-related topics. 
}

Our DS curriculum builds on the following units at FCFM:
\begin{itemize}
    \item The Center for Mathematical Modelling (CMM). Areas: probability, optimisation, statistical machine learning, machine learning for healthcare, and scientific  computing.
    \item The Department of Computer Science (DCS). Areas: data mining, natural language processing, database theory, deep learning, multimedia dabatabases, information retrieval, semantic web, and data compression.
    \item The Department of Electrical Engineering (DEE). Areas: signal processing, computational intelligence, robotics, information theory, and control systems. 
\end{itemize}


\subsection{Learning objectives} 
\label{sub:philosophy}

\begin{figure}[h]
\centering
\includegraphics[width=0.45\textwidth]{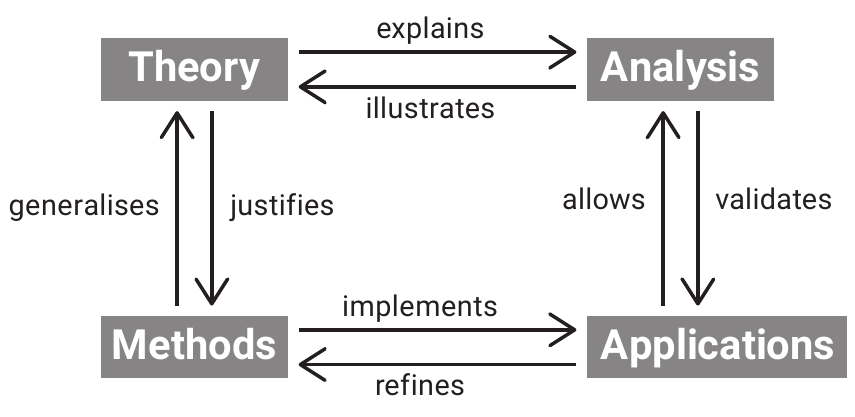}
\caption{Learning objectives and their connections: Theory supports the development of Methods and ways of generalising them to different scenarios. At the same time, Applications provide insight into the choice and enhancement of Methods, while the results of Applications allow for Analysis which validates or refutes the Theory on which the other learning objectives are based.}
\label{fig:diagram}
\end{figure}

\revised{We integrate the engineering perspective into teaching DS by pursuing four learning objectives. 
The first, rooted in \textbf{Theory}, relates to understanding the data-generating systems to identify challenges and envision solutions at a conceptual level. This can be achieved \emph{from first principles} or from a data-driven, application-agnostic, machine learning perspective. 
The second objective relates to selecting the \textbf{Methods} for the different stages of DS. This is fundamental for practitioners who should be able to discriminate which tools are appropriate for each task, both at the level of data handling and knowledge extraction. 
The third objective relates to \textbf{Applications}, through which professionals deal with real-world DS challenges of different natures by formulating the problem in a DS setting where attaining a solution is feasible. The final learning objective focuses on \textbf{Analysis}, in which professionals are expected to interpret the results so as to i) provide explanations, ii) identify possible shortcomings of the methods employed, and iii) explore solutions for such shortcomings based on theory. 

Our learning objectives are interconnected and they support one another, as illustrated in the diagram in Fig \ref{fig:diagram}. Lastly, it is worth noting that our objectives are in line with other criteria for accreditation of engineering programmes such as those of the UK's Accreditation of Higher Education Programmes (AHEP), set by the Engineering Council \cite[p.10]{eng_council} and also by the The Institution of Engineering and Technology (IET) \cite[p.8]{iet}.}

\section{Components of the DS Ecosystem}
\label{sec:components}

\subsection{Master of Science programmes}
\label{ssec:programmes}

The bases of our DS ecosystem are the MSc programmes at the units mentioned in Sec.~\ref{sub:our_units}: the MSc in Mathematical Modelling, MSc in Computer Science and MSc in Electrical Engineering, all of which feature a DS specialisation \revised{and are accredited in the country by the corresponding institutions}. Hosted at different departments, these programmes offer complementary views, where students are exposed to courses, students and faculty of different departments. This collaborative environment allows our students to build their own DS profile, by mixing resources from different perspectives. 

\subsection{Postgraduate courses}
\label{ssec:courses}

\begin{table*}[hbt]
\centering
\caption{Courses in the DS ecosystem grouped per unit as described in Sec.~\ref{sub:our_units}. The number of students corresponds to the last term the course was given. \revised{The course level can be identified from its code: 4000 or less corresponds to undergraduate courses, 5000 corresponds to both undergraduate and postgraduate, and 6000 courses and above are restricted to postgraduate students.}}

\resizebox{\textwidth}{!}{
\begin{tabular}{|l|c|c|c|c|c|c|c|c|c|c|} \hline \hline
\multicolumn{4}{|c|}{} & \multicolumn{3}{c|}{ Content } & \multicolumn{4}{c|}{ Focus } \\ \hline
course & code & students/term & department & Theory & Methods & Applications & D. Management & D. Analytics & A. Domain & R. Field \\ \hline \rowcolor{Gray}
Statistics & MA3402 & 48 & CMM & \checkmark & \checkmark & $\cdot$ & $\cdot$ & \checkmark & $\cdot$ & $\cdot$ \\ 
Stochastic Simulation & MA4402 & 29 & CMM & \checkmark & $\cdot$ & $\cdot$ & $\cdot$ & $\cdot$ & $\cdot$ & \checkmark \\ \rowcolor{Gray}
Machine Learning & MA5204 & 90 & CMM & \checkmark & \checkmark & $\cdot$ & $\cdot$ & \checkmark & $\cdot$ & $\cdot$ \\ 
Advanced Machine Learning & MA5309 & 7 & CMM & \checkmark & \checkmark & $\cdot$ & $\cdot$ & \checkmark & $\cdot$ & $\cdot$ \\ \rowcolor{Gray}
Probability \& Statistics for Data Science & MA5406 & 12 & CMM & \checkmark & \checkmark & $\cdot$ & $\cdot$ & \checkmark & $\cdot$ & $\cdot$ \\
Lab of Mathematical Modeling & MA5500 & 7 & CMM & $\cdot$ & $\cdot$ & \checkmark & $\cdot$ & $\cdot$ & $\cdot$ & \checkmark \\  \rowcolor{Gray}
Optimisation for Data Science & MA5705 & 21 & CMM & \checkmark & \checkmark & $\cdot$ & $\cdot$ & $\cdot$ & $\cdot$ & \checkmark \\
Scientific Computing & MA6201 & 8 & CMM & $\cdot$ & $\cdot$ & \checkmark & \checkmark & \checkmark & $\cdot$ & $\cdot$ \\  \rowcolor{Gray}
Data Science Laboratory & MA6202 & 27 & CMM & $\cdot$ & \checkmark & \checkmark & \checkmark & \checkmark & $\cdot$ & $\cdot$ \\ \hline
Algorithms and Data Structures & CC3001 & 186 & DCS & \checkmark & \checkmark & $\cdot$ & $\cdot$ & $\cdot$ & $\cdot$ & \checkmark \\ \rowcolor{Gray}
Databases & CC3201 & 117 & DCS & \checkmark & \checkmark & $\cdot$ & \checkmark & $\cdot$ & $\cdot$ & $\cdot$ \\
Introduction to Data Mining & CC5206 & 70 & DCS & $\cdot$ & \checkmark & \checkmark & \checkmark & \checkmark & $\cdot$ & $\cdot$ \\ \rowcolor{Gray}
Information Visualization & CC5208 & 17 & DCS & $\cdot$ & $\cdot$ & \checkmark & $\cdot$ & \checkmark & $\cdot$ & $\cdot$ \\
Massive Data Processing & CC5212 & 60 & DCS & $\cdot$ & \checkmark & \checkmark & \checkmark & $\cdot$ & $\cdot$ & $\cdot$ \\ \rowcolor{Gray}
Data Science Project & CC5214 & 17 & DCS & $\cdot$ & $\cdot$ & \checkmark & \checkmark & \checkmark & $\cdot$ & $\cdot$ \\
Image Processing and Analysis & CC5508 & 14 & DCS & $\cdot$ & \checkmark & $\cdot$ & $\cdot$ & \checkmark & \checkmark & $\cdot$ \\  \rowcolor{Gray}
Pattern Recognition & CC5509 & 11 & DCS & $\cdot$ & \checkmark & $\cdot$ & $\cdot$ & \checkmark & $\cdot$ & $\cdot$ \\
Business Analytics & CC5615 & 38 & DCS & $\cdot$ & \checkmark & \checkmark & \checkmark & \checkmark & $\cdot$ & $\cdot$ \\   \rowcolor{Gray}
Deep Learning & CC6204 & 118 & DCS & \checkmark & \checkmark & $\cdot$ & $\cdot$ & \checkmark & $\cdot$ & $\cdot$ \\
Natural language processing & CC6205 & 57 & DCS & $\cdot$ & \checkmark & \checkmark & $\cdot$ & \checkmark & \checkmark & $\cdot$ \\ \rowcolor{Gray}
Web of Data & CC7220 & 36 & DCS & $\cdot$ & \checkmark & \checkmark & \checkmark & $\cdot$ & $\cdot$ & $\cdot$ \\  \hline 
Signal Processing & EL4101 & 11 & DEE & \checkmark & \checkmark & $\cdot$ & $\cdot$ & \checkmark & \checkmark & $\cdot$ \\
\rowcolor{Gray}
Computational Intelligence & EL4106 & 51 & DEE & \checkmark & \checkmark & $\cdot$ & $\cdot$ & \checkmark & $\cdot$ & \checkmark \\
Neural Networks and Information Theoretic Learning & EL7006 & 11 & DEE & \checkmark & \checkmark & $\cdot$ & $\cdot$ & \checkmark & $\cdot$ & $\cdot$ \\ \rowcolor{Gray}
Introduction to Digital Image Processing & EL7007 & 16 & DEE & $\cdot$ & \checkmark & $\cdot$ & $\cdot$ & \checkmark & \checkmark & $\cdot$ \\
Advanced Image Processing & EL7008 & 36 & DEE & \checkmark & \checkmark & $\cdot$ & $\cdot$ & \checkmark & \checkmark & $\cdot$ \\ \rowcolor{Gray}
Fault Diagnosis and Failure Prognosis & EL7014 & 19 & DEE & $\cdot$ & \checkmark & $\cdot$ & $\cdot$ & \checkmark & \checkmark & $\cdot$ \\
Robotics, Sensing and Autonomous Systems & EL7021 & 4 & DEE & $\cdot$ & \checkmark & $\cdot$ & $\cdot$ & \checkmark & \checkmark & \checkmark \\ \rowcolor{Gray}
Information Theory: Fundamentals and Apps. & EL7024 & 26 & DEE & \checkmark & $\cdot$ & $\cdot$ & $\cdot$ & $\cdot$ & $\cdot$ & \checkmark \\ \hline \hline
\end{tabular}
} \vspace{1mm}

\label{tab:courses}
\end{table*}

We describe DS-related courses using a two-level categorisation, where categories (\emph{Content} and \emph{Focus}) and subcategories are not necessarily mutually exclusive. \revised{The \emph{Content} category points to the elements that are taught in each course and follows from our first three learning objectives in Sec.~\ref{sub:philosophy}. The \emph{Focus} category relates to the DS stages towards which each course aims.} The complete list of courses can be found in Table \ref{tab:courses}.

\subsubsection{Content} \textbf{Theory} courses are oriented to the formulation and analysis of mathematical and computational models for data management and data analysis. Theory-based courses enable students to understand the limitations of off-the-shelf methods and to question existing solutions. Topics for such courses include probability, statistics, stochastic processes, optimisation, algorithmic complexity, discrete mathematics, database theory, and dynamical systems.

\textbf{Methods} courses focus on addressing practical DS challenges in a problem-driven fashion. The core of DS methods includes natural language processing (NLP), deep learning (DL), signal processing (SP), non-parametric Bayesian inference, transform-based analysis, spectral analysis, Monte Carlo simulation, and data visualisation. 

Lastly, \textbf{Applications} courses ensure that students are not only knowledgeable on theory and methods but can also implement them on  arbitrary-domain challenges. This is particularly useful for graduates working in a \emph{data science as a service} environment such as  business analytics, health, climate or astronomy. To meet the wide range of student interests and needs, we place particular attention on recreating realistic DS scenarios within our courses so that students face all stages of a real-world DS project, i.e., from data management (acquisition, curation and processing) to data analytics (mining, inference, decision making and interpretation). 

\subsubsection{Focus} A course can focus on one or more of the following aspects.
\begin{itemize}
\item[a)] \textbf{Data Management:} topics of data handling such as acquisition, processing, governance, architecture,  storage, security, privacy, quality and curation.
\item[b)] \textbf{Data Analytics:} knowledge extraction tools such as  machine learning, probability models, statistics, time series, and data mining. 
\item[c)] \textbf{Application Domains:} areas that rely on DS resources such as text processing, speech synthesis, computer vision, image processing, robotics, econometrics, astro-statistics, bio-informatics,  among others.
\item[d)] \textbf{Related Fields:} disciplines, not necessarily associated with DS, where knowledge extraction is also relevant (information theory, stochastic simulation, ergodic theory and dynamical systems, and algorithms) or  disciplines that provide skills necessary for data scientists (optimisation, algebra, algorithms, and stochastic processes).
\end{itemize}

\begin{figure}
\centering
\includegraphics[width=0.49\textwidth]{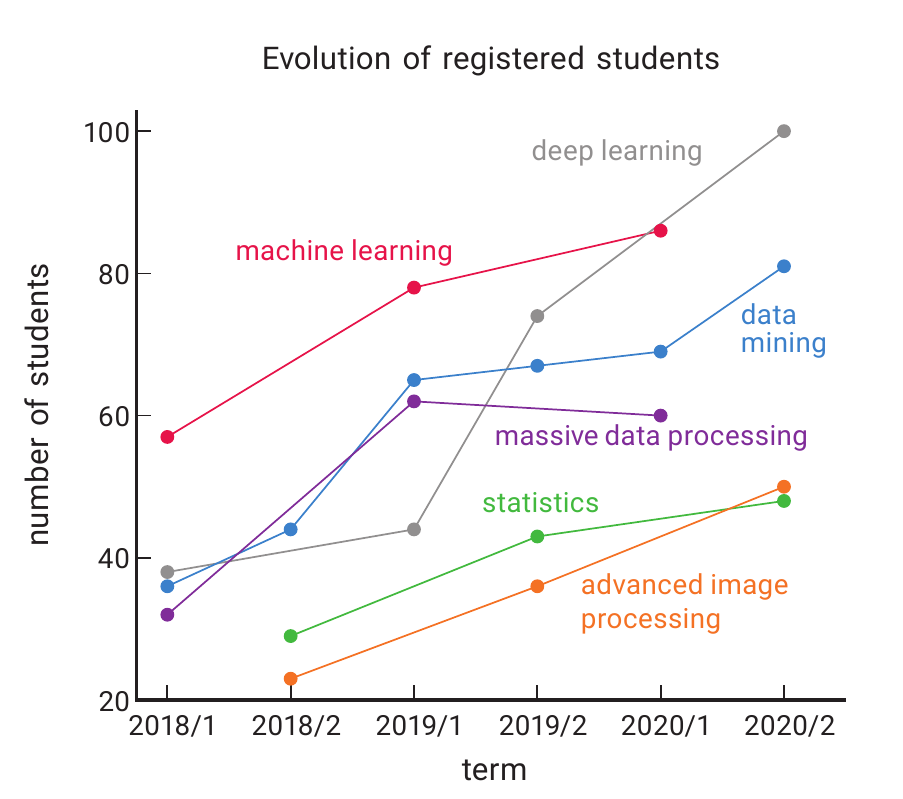}
\caption{Evolution of registered student in the most relevant DS courses over the last three years.} 
\label{fig:courses_plot}
\end{figure}

\revised{Though these courses are offered as a part of the aforementioned MSc programmes, they are available for all the students at FCFM, provided they meet course requirements. Additionally, as part of different programmes, some course content may overlap (e.g., machine learning and computational intelligence); however, despite this redundancy, the courses have gained considerable popularity of late.  Fig.~\ref{fig:courses_plot} shows that the number of students has increased over the last three years for the flagship courses. Lastly, it is worth mentioning that the contents of these courses have evolved and will continue to evolve over time based on feedback from students and on the current state of the art in the field. For some of the courses, the content is publicly available, such as those in Table \ref{tab:courses_website}.}

\subsection{Internships} 
\label{ssec:internships} 

Our DS offerings include internships oriented to professional or research work. Those in professional internships, i.e., interns working on our projects or those of our collaborators, can delve deeper into the practice of DS and thus make informed decisions when choosing a DS career.

Research internships provide a unique opportunity for students in the transition undergrad, MSc or PhD programmes, students interested in a temporary (usually summer) position, and even visiting students joining mainly from our partner institutions. Research internships provide students with first-hand experience in a DS research laboratory. Funding for research interns (both domestic and international) has been possible through faculty research funds, collaboration networks, and also internationalisation grants.

\section{Innovative aspects of our approach}
\label{sec:innovations}


Though it is the practical advantages of DS what usually motivates students to seek training in the field, a necessary step to become proficient in DS is to understand the required theory. As pointed out in \cite[Sec.~2.5]{curr_guide_2017}, however, theory cannot be delivered in a raw manner for DS students as in classical scientific degrees, but instead it should be presented in a problem-driven fashion. Additionally, even for those students who are familiar with the theory already, e.g., those holding a degree in mathematics, making the transition from theory to DS practice can be a challenge. 
Current information technologies, computational resources, and public datasets allow us to offer an \textit{ad hoc}, pedagogical, presentation of the theory and its connection to DS practice. These innovations are key in our ecosystem, just as they have been for teaching statistical signal processing \cite{sithan} or AI \cite{Kumar98}. 

We next describe the innovative features in our DS ecosystem (either exploratory or consolidated) and how they enrich the pedagogical process. 

\subsection{Online code repository} As a companion to some of our courses, we include the pedagogical material in a public repository. This allows students to have instant access to lecture notes, slides, assignments, demonstrations and, in some cases, videolectures. This way of distributing the material has proven advantageous for several reasons. First, both the lecturer and the teaching assistants can simultaneously edit the material, minimising the amount of conflicts and maintaining a history of past versions. Second, should last-minute changes occur in the course materials, the up-to-date version is automatically available to the students. Third, the students can visualise the course contents online, which is of particular interest for Jupyter notebooks (see next section) that require a specific interpreter in local machines. Fourth, all course material can be made universally available to the general public beyond our institution. \revised{See Table \ref{tab:courses_website} for an example of our courses GitHub repositories.}

\begin{table}[t]
\centering
\caption{GitHub repositories (hyperlinks) of some of our courses containing lecture notes, slides, exercises and demos.}
\label{tab:courses_website}
\begin{tabular}{|l|p{5.7cm}|}
\hline
\textbf{Course}             & \textbf{GitHub address (append https://github.com/)} \\ \hline
Machine learning            & \href{https://github.com/GAMES-UChile/Curso-Aprendizaje-de-Maquinas}{GAMES-UChile/Curso-Aprendizaje-de-Maquinas}  \\ \hline
Statistics                  & \href{https://github.com/GAMES-UChile/Curso-Estadistica}{GAMES-UChile/Curso-Estadistica} 
                                    \\ \hline
Deep learning               &
\href{https://github.com/dccuchile/CC6204}{dccuchile/CC6204}                                                   \\ \hline
Natural lang.~proc. &
\href{https://github.com/dccuchile/CC6205/}{dccuchile/CC6205}                                                 \\ \hline
\end{tabular}
\end{table}

\subsection{Interactive programming} \label{sub:jupyter} Computer programming is best taught with a \emph{learning-by-doing} approach \cite{reese2011learning}. For DS in particular, the Jupyter notebook (JN) has revolutionised the way we programme \cite{perkel2018jupyter}, with a clear impact when teaching and prototyping: it is free, open-source, interactive, intuitive, and supported by a strong online community. Our courses feature programming modules on Python (for a majority of courses), R (statistics), MATLAB (elec.~eng.), and C++ (scientific computing). Therefore, as JNs are compatible with all these languages, they are used for in-class demonstrations, in which the lecturer can produce and run examples on the fly. Additionally, these JNs are distributed to the students for personal study, allowing them to complete, modify and run examples at their convenience while also exploring creative variants; this is especially required for assignments and project-oriented activities within the courses. Beyond methods-based courses, where the value of code demonstrations is clear, we have learnt that JNs and similar software constitute an excellent complement for theory-based courses too. For example, illustrations of hypothesis testing in statistics and stochastic gradient descent in optimisation greatly benefit from modifiable demonstrations when compared to old-school blackboard illustrations.

\subsection{Evaluations promoting independence and creativity} To a large degree, course evaluations condition the design of the course and its success in transmitting knowledge. A successful evaluation becomes particularly challenging in the context of our teaching objectives (outlined in Section \ref{sub:philosophy}), which aim at having students  equally comfortable with both theory and practice. As DS challenges require creative, out-of-the-box solutions, we strive to recreate these requirements in our evaluations. Whenever the topics allow it, in addition to the theoretical/practical parts, our evaluations incorporate open-ended questions whose objective is to encourage students to build on the concepts learnt in the lectures. In these instances, students are required to specify the question and solve it, derive alternative solutions to those problems examined in class, or review the literature for material that has been hinted at in class yet not thoroughly reviewed. In this way, we aim to ensure that students solve realistic problems rather than (just) implementing an off-the-shelf method. This has proven to be particularly challenging for inexperienced students used to well-defined problems, which often have a unique solution; these students require close supervision. 

 \subsection{Project-oriented learning}  In most of our courses, the final evaluation requires students to form groups (of 2-4 members) to complete a project, which can be of theoretical or applied content, or a combination of both. The execution of such a project is developed throughout the course, alongside lectures, where preliminary advances of the projects are monitored as partial course evaluations (tests and assignments). We usually provide students with a repository of project topics built from past courses, industrial projects, and the lecturers' own research portfolio. However, the students are also encouraged to propose project themes motivated by their thesis work, entrepreneurial activities, or other topics of interest that can be addressed using DS. We have noticed that it is precisely those projects brought by the students that turn out to be the most successful, most likely because there is a genuine motivation to work on these rather than on a generic assignment. 

\subsection{Communication skills} DS engineers work in interdisciplinary teams and must be able to communicate clearly. Our courses consider four practices aimed at developing these skills. First, the format for the submissions (assignments and reports) is evaluated in terms of presentation, conciseness, clarity and readability. What students have found particularly challenging here is to constrain their description to a limited number of pages. Second, in seminar-based courses, we use the \emph{flipped-classroom} method \cite{bishop2013flipped}, where students teach their fellow classmates. Third, in project-oriented courses, students work on a DS problem for which they must formulate and define the scope, select the methods and strategies to be used, and analyse the results. In all these stages, the students work as a team: regular meetings and presentations are conducted to evaluate the ability of the students to communicate the project's state to the rest of the class and the instructor. Fourth, for those courses featuring real-world projects from industry, students present their (finalised) DS projects in a 10-minute pitch talk to the company that proposed the challenge, thus validating the communication abilities of the students with actual industrial counterparts.

\subsection{Early research training}
\label{ssec:early_research}

The MSc programmes described in Section \ref{ssec:programmes} culminate with a two-semester research thesis that can be of either an applied or theoretical nature; for most students, this experience constitutes their first exposure to research. In their theses, students join other research students and one or more supervisors, sometimes in collaboration with partners from the industry, the public sector or other sciences. \revised{Exposing our students to research may also improve their employability: though the majority of our graduates join the industrial sector, their research experience makes them valuable assets in modern industry, which often values research. }

Lastly, for those of our graduates who pursue an academic career, the MSc thesis provides a fertile environment for theoretical research, where students join PhD students and postdocs to work under the close supervision of their mentors and, in most cases, successfully publish their findings (see, e.g., \cite{MSc1,MSc2,claveria-2019,BarceloM0S20,dunstan2020predicting}).

\section{Professional and Academic Development}
\label{sec:for_professionals}

The public and private (aka \emph{professional}) sectors as well as academia have witnessed the practical advantages of DS and are eager to understand and incorporate such techniques. We have addressed the demand of these sectors for DS training by transferring our experience from undergraduate and graduate programmes at FCFM to the professional domain. We next describe the elements of our DS training focused on professionals.

\subsection{Professional diplomas} \label{ssub:diplomas} Continuing education courses are the most popular destination for professionals seeking DS training. \revised{To stand out from the abundance of offerings from other institutions, the distinguishing feature of our professional diplomas follows from our learning objectives and Master's programmes (Secs.~\ref{sub:philosophy} and \ref{ssec:programmes}) to provide an alternative that blends theory and practice in a problem-oriented manner.} In particular, we offer two diplomas relevant to DS through the Department of Computer Science (Section \ref{sub:our_units}): the \emph{Data Science diploma} and the more advanced \emph{Artificial Intelligence diploma}. Each of these diploma programmes features three evening lectures per week, which are completed over a five-month period.

Both diplomas target professionals from the areas of Engineering and Science, such as astronomers, geologists, biologists, and engineers, although sociologists and lawyers have also successfully completed the courses. \revised{On one hand, the Data Science diploma focuses on the analysis and handling of complex and massive datasets; the main topics are those related to fundamentals of databases and data mining, basic statistical tools, big data, information retrieval, and visualisation. The Artificial Intelligence diploma, on the other hand, focuses on a more experienced audience (e.g., those graduated from the previous diploma), to train them to i) lead projects that involve complex and heterogeneous data sources in various forms (e.g., text, images) and ii) effectively communicate and justify their findings. Accordingly, this second diploma features more specific contents such as deep learning, evolutionary algorithms, image processing, natural language processing, and robotics. Lastly, both diplomas feature a final project, through which students tackle a challenge relevant to their own  workplace under the supervision of an academic staff member.}

\subsection{On-demand courses} We have also developed tailored courses for those in the professional sector who currently work in DS. These courses have been offered through the Center for Mathematical Modeling (see Sec.~\ref{sub:our_units}) to partners in banking, mining, and NGOs that aim to acquire specific and advanced DS skills. For these courses, the syllabus is jointly designed with the interested party with their particular needs and challenges in mind. The courses work as a blend between a diploma and a scientific consultancy, whereby the class demonstrations utilise data provided by the institution. In this way, students learn the impact and shortcomings of standard methods, as well as the necessity for developing new tools in a familiar environment and with a clear (problem-driven) purpose. 

As a consequence of the interdisciplinarity of DS teams in industry, a recurrent challenge in these on-demand courses comes from the heterogeneous levels of expertise in DS found among students in the same group. This justifies the development of courses for small groups of students with purpose-specific content. In fact, when done face-to-face, we have found that groups of approximately 15 students in weekly sessions of 2-3 hours (with a break) are an appropriate format. This allows us to assess the evolution of the students via discussions in class rather than relying on strict evaluations, which are usually incompatible with the availability of the students in these programmes. Additionally, these courses employ most of the innovations described in Sec.~\ref{sec:innovations}, especially those related to the demonstrations using JNs and Github repositories. 

Due to the coronavirus outbreak and the sustained lockdown measures, during 2020 we have offered our tailored courses in an online format. These have been particularly useful for mining companies, where engineers both in Santiago and close to the extraction sites have taken part in the courses. When implemented remotely, we have combined online classes and offline content capsules for the students to manage at their convenience.  

\subsection{Training of data engineers}
\label{ssec:trainee}

The training of DS engineers involves a wide range of coursework, experiences, and exposure to different sectors and professionals both within academia and industry. The data engineers who go through our programmes have the opportunity to work on our research projects alongside the principal investigators, research assistants, postgraduate students and interns; for technology transfer projects, the teams usually comprise engineers, (data) analysts, designers and software developers. In our collaborative teams, project engineers are constantly exposed to research practice and, depending on the nature of the project, they even participate in applied research publications (see, e.g., \cite{DEWOLFF2020,delplancke2020scalable,baez2020chilean}). Furthermore, many of the engineers who take part in our programmes will have the opportunity to teach and train others, which further enriches their own development and keeps them current on the latest trends and advances in the field.   

Within our laboratories and centers, providing hands-on training in which we tackle problems from different industries not only provides an invaluable DS experience for our project engineers, but it also reinforces our role as educators, bridging the gap between academia and the professional sector. Additionally, when our former engineers move on to find jobs outside  academia, our collaboration network is strengthened and our new project pipeline is also often extended. 

Indeed, the considerable demand for DS professionals makes the DS job market quite dynamic: project engineers move from academia to industry and also between companies perhaps more than in other disciplines. One reason for this is that DS projects are often completed on a contract basis in which engineers are hired for a specific time to solve a specific problem. This practice is supported by the perspective that sharing talent across different sectors and job mobility are regarded as positive for career development in the field of DS \cite{the2019dynamics}.  This job mobility further underscores the need for our programmes to provide a wide exposure to the types of problems and experiences DS engineers are likely to encounter once they leave our centres.

\section{Outreach}
\label{sec:outreach}

In addition to the formal treatment of DS in academia, the private and public sectors, we have worked towards making DS advances available to the general public. We consider this to be part of the role of universities in the \emph{democratisation} of knowledge \cite{biesta2007towards} which, in our case, relates to promoting literacy in DS. This can be achieved by scientific dissemination activities organised by the public sector or NGOs. In particular, owing to the contemporary online teaching practices, outreach also needs to occur through (virtual) talks and discussion panels and \emph{webinars}, which can be backed up on a video (e.g., YouTube) repository site.

Furthermore, to raise awareness about the impact of DS on industry leaders and policy makers, we have held the “Data-Days”, a series of discussion panels organised since 2018 where participants discuss a particular DS topic with an influential invited expert. Topics considered so far have been clinical text mining, digitalisation of education, climate change and biodiversity, social organisation and representativeness, and the Internet of Things in healthcare. These panels are designed to encourage discussion between attendees and experts so as to identify opportunities and challenges related to modernisation of the local economy via DS. 

Another initiative for disseminating the advances and impact of DS is through seminars addressed to high-school and university students. 
For secondary students in particular, we have seen that DS and AI are becoming popular. In this sense, the Explora outreach programme\footnote{Driven by the Chile's Ministry of Education, see \url{https://www.explora.cl/}.} invites secondary students to develop a project under the supervision of a DS expert. In fact, some high schools have instituted (Python) programming courses through which our researchers have carried on a vibrant interaction. Finally, it is relevant to mention that there are initiatives that aim to reduce the gender gap in STEM disciplines, and many of these events have recruited DS experts to give open talks or serve as judges in DS competitions.

\section{Open challenges}
\label{ssec:open_c}

Teaching DS focuses on shaping highly-skilled technical professionals to develop and implement methods to extract information from data in various domains. However, being in close connection to AI, the discipline of DS is also at risk of being automated itself. Therefore, the following question arises naturally: How should we cope with the replacement of data scientists by machines? It is known that Google, Amazon and IBM do provide cost-efficient, modular, DS solutions that are the choice of companies relying on DS \emph{as a service} (DSAAS); it is critical that our graduates can deal with and adapt to the massification of DSAAS. To this end, our graduates should master the underlying theory of DS practices so they truly are data \emph{scientists}, and not mere data science \emph{practitioners}, and can adapt to changing circumstances in their field of expertise. 

Another challenge to be faced by our graduates is that of the so-called \emph{social value of the data}. Novel tools for data processing have allowed us to identify their value as a means to multiple ends such as marketing, political campaigns, public policies and insurance. There are, of course, companies that support their activity purely on the value of data, such as Twitter and Facebook. With the sophistication of DS tools to extract information from data, we are facing an era where---to an extent---data can be considered a commodity; this scenario opens both negative and positive opportunities. First, how can we guarantee that as a small country we are able to protect our data, when large international conglomerates are at play? For instance, Chile's recently-launched Data Observatory\footnote{\url{https://www.dataobservatory.net/}} will be hosted at Amazon Web Services, which has implications unknown to the public at the time of this writing. Second, are we able to take the leap forward into a modern, technological society by both properly curating our data and developing tools to extract knowledge from them? There are case studies of which our students should be aware in this regard, such as that of Cambridge Analytica \cite{cambridge_analytica}. As a consequence, our DS curriculum should feature courses dedicated to the issue of data value and privacy so that our graduates, in addition to being experts on deep learning, scientific computing and probabilistic modeling, are also knowledgeable of the value and impact conveyed by the DS tools they handle.

\section{Summary}
\label{sec:summary}

Developing the described teaching ecosystem has been an enriching experience both as researchers and educators. Through this article, we have highlighted the  considerations and innovative aspects we  consider meaningful and essential in putting together effective DS curricula for undergraduate and postgraduate students, professionals and the general public. There is a common denominator in designing a DS ecosystem: finding the appropriate balance among theory, methods, and applications. This interplay is essential to achieve an educational experience for the students that is practical (an up-to-date presentation of techniques and solutions), meaningful (covering the advantages and the limitations of the strategies and methods), and fundamental (promoting critical thinking and a level of abstraction that facilitate innovation and creativity in DS). Considering our DS as an ecosystem rather than as a single curriculum has allowed us to widen our scope and include not just courses but resources and outreach initiatives. We hope that the material presented here can help others in the process of developing their own DS programmes and ecosystems.

\section*{Acknowledgements}

We are grateful to the Faculty of  Physical  and  Mathematical  Sciences (FCFM),  Universidad  de  Chile, for its continuous support of DS initiatives. In particular, we deeply thank our colleagues at the Dept.~of Electrical Engineering, the Dept.~of Computer Science and the Center for Mathematical Modeling, as they have made possible the DS ecosystem  described here.

This work was supported by the following ANID grants: AFB170001 Center for Mathematical Modeling, AFB0008 Advanced Center of Electrical and Electronic Engineering, ICN17\_002 Millennium Institute for Foundational Research on Data, and Fondecyt-Iniciación projects \#11171165, \#11200290 \& \#11201250.

\bibliography{biblio}{}
\bibliographystyle{IEEEtran}
\end{document}